\documentclass[twocolumn,pre,showpacs]{revtex4}
\usepackage{amsmath,amsthm}
\usepackage{graphicx}
\usepackage{dcolumn}
\usepackage{bm}
\usepackage{float} 

\begin{document}

\title{Maximally informative pairwise interactions in networks}

\author{Jeffrey D. Fitzgerald and Tatyana O. Sharpee} 

\affiliation{Computational Neurobiology Laboratory, The Salk Institute for Biological Studies, La Jolla, CA 92037, USA \\ The Center for Theoretical Biological Physics, University of California-San Diego, La Jolla, California 92093, USA}

\begin{abstract}
Several types of biological networks have recently been shown to be
accurately described by a maximum entropy model with pairwise
interactions, also known as the Ising model.  Here we present an
approach for finding the optimal mappings between input signals and
network states that allow the network to convey the maximal
information about input signals drawn from a given distribution.  This
mapping also produces a set of linear equations for calculating the
optimal Ising model coupling constants, as well as geometric properties that 
indicate the applicability of the pairwise Ising model.  We show that the optimal
pairwise interactions are on average zero for Gaussian and uniformly
distributed inputs, whereas they are non-zero for inputs approximating
those in natural environments.  These non-zero network interactions
are predicted to increase in strength as the noise in the response
functions of each network node increases.  This approach also suggests
ways for how interactions with unmeasured parts of the network can be
inferred from the parameters of response functions for the measured
network nodes.

\end{abstract}

\date{\today}
\maketitle

\section{Introduction}
Many organisms rely on complex biological networks both within and between cells to process information about their environments \cite{Bray_1995,Network_complexity}.  As such, their performance can be quantified using the tools of information theory \cite{Cover_Thomas,Tcacik_etal_PRE_2008,Tostevin_2008,Ziv_etal_2008}. Because these networks often involve large numbers of nodes, one might fear that difficult-to-measure high-order interactions are important for their function. Surprisingly, recent studies have shown that neural networks \cite{Schneidman_2006,Shlens_2006,Tang_Beggs_2008,Pillow_2008}, gene regulatory networks \cite{Federoff_2006,Wolynes_2009}, and protein sequences  \cite{Socolich_2005,Bialek_Ranganathan_2007} can be accurately described by a maximum entropy model including only up to second-order interactions. In these studies the nodes of biological networks are approximated as existing in one of a finite number of discrete states at any given time. In a gene regulatory network the individual genes are binary variables, being either in the inactivated or the metabolically expensive activated states.  Similarly, in a protein the nodes are the amino acid sites on a chain which can take on any one of twenty values.  

We will work in the context of neural networks, where the neurons communicate by firing voltage pulses commonly referred to as "spikes" \cite{Rieke_book}.  When considered in small enough time windows, the state of a network of $N$ neurons can be represented by a binary word $\sigma=\left(\sigma_{1}, \sigma_{2},\ldots,\sigma_{N} \right)$, where the state of neuron \textit{i} is given by $\sigma_{i}=1$ if it is spiking and $\sigma_{i}=-1$ if it is silent, similar to the $\uparrow$/$\downarrow$ states of Ising spins.  

The Ising model, developed in statistical physics to describe pairwise interactions between spins, can also be used to describe the state probabilities $P_{\sigma}$ of a neural network:
\begin{equation}
\label{Ising Model Eq}
P_{\sigma} = \dfrac{1}{Z} \textrm{exp} \left[ \displaystyle\sum_{i} h_{i} \sigma_{i} + \displaystyle\sum_{i \neq j} J_{ij} \sigma_{i} \sigma_{j}\right].
\end{equation}\\
Here, $ Z $ is the partition function and the parameters $\left\lbrace h_{i}\right\rbrace $ and $ \left\lbrace J_{ij}\right\rbrace$ are the coupling constants. This is the least structured (or equivalently, the maximum entropy \cite{Jaynes_1957}) model consistent with given first- and second-order correlations, obtained by measuring $ \left\langle \sigma_{i} \right\rangle $ and $ \left\langle \sigma_{i} \sigma_{j} \right\rangle $, where these averages are over the distribution of network states.

In magnetic systems one seeks the response probabilities from the coupling constants, but in the case of neural networks one seeks to solve the inverse problem of determining the coupling constants from measurements of the state probabilities. Because this model provides a concise and accurate description of response patterns in networks of real neurons \cite{Schneidman_2006,Shlens_2006,Tang_Beggs_2008,Pillow_2008}, we are interested in finding the values of the coupling constants which allow neural responses to convey the maximum amount of information about input signals.

The Shannon mutual information can be written as the difference between the so-called response and noise entropies \cite{Cover_Thomas}: 
\begin{equation}
\label{info}
I=H_{\rm{resp}}-H_{\rm{noise}}. \nonumber
\end{equation}
The response entropy quantifies the diversity of network responses
across all possible input signals $\{ \mathcal{I} \} $. For our
discrete neural system this is given by
\begin{equation}
\label{Hresponse}
H_{\rm{resp}} = - \displaystyle\sum_{ \left\lbrace \sigma \right\rbrace } 
P_{\sigma} \log_2 \left( P_{\sigma} \right).
\end{equation}
In the absence of any constraints on the neural responses,
$H_{\rm{resp}}$ is maximized when all $2^{N}$ states are equally
likely \cite{Attwell_Laughlin_2001}.

The noise entropy takes into account that the network states may vary
with respect to repeated presentations of inputs, which reduces the
amount of information transmitted.  The noise entropy is obtained by
computing the conditional response entropy $ P_{\sigma|{\mathcal{I}}}
$, and averaging over all inputs,
\begin{equation}
\label{Hnoise}
H_{\rm{noise}} = -\displaystyle\int d\mathcal{I}~ P_{\mathcal{I}} \displaystyle\sum_{ \left\lbrace \sigma
\right\rbrace }  P_{\sigma|{\mathcal{I}}} \log_2
\left( P_{\sigma|{\mathcal{I}}} \right),
\end{equation}
where $P_{\mathcal{I}}$ is the input probability distribution. Thus in
order to find the maximally informative coupling constants, we must
first confront the difficult problem of finding the optimal mapping
between inputs $\mathcal{I}$ and network states $\sigma$.

\section{Decision boundaries}
The simplest mappings from inputs to neural response involve only a single input dimension \cite{deBoer,Meister99,Schwartz06,Victor80}.  In such cases, the response of a single neuron can often be described by a sigmoidal function with respect to the relevant input dimension \cite{Rieke_book,Laughlin_1981}.  However, studies in various sensory systems, including the retina \cite{Fairhall_2006}, the primary visual \cite{Rust05,Chen07,Touryan02,Felsen05}, auditory \cite{Atencio_2008}, and  somatosensory \cite{Maravall_etal_2007} cortices, have shown that neural response can be affected by multiple input components, resulting in highly nonlinear, multi-dimensional mappings from input signals to neural responses. 

\begin{figure}[t]
\centering
\includegraphics[width=80mm]{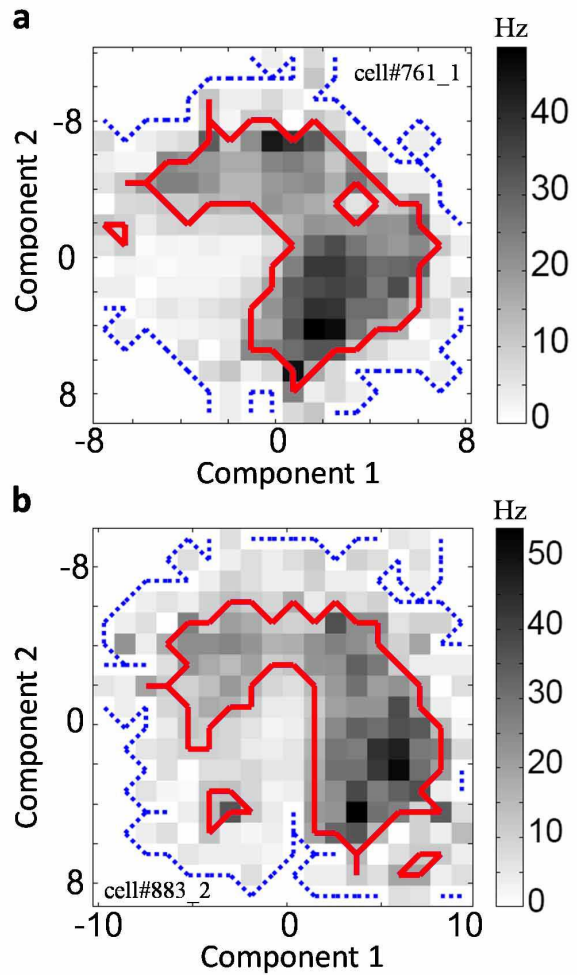}
\caption{(Color online) Example analysis of firing rate for a simple, \textbf{a}, and
a complex, \textbf{b}, cell in the cat primary visual cortex probed
with natural stimuli from the dataset \cite{Sharpee06}. Two relevant input dimensions were found for
each neuron using the maximally informative dimensions method
described in \cite{Sharpee04}.  Color shows the firing rate as a
function of input similarity to the first (x-axis) and second (y-axis)
relevant input dimensions. The values on the x- and y- axis have been
normalized to have zero mean and unit variance. Blue (dashed) lines show
regions with signal-to-noise ratio $>$ 2.0. Red (solid) lines are drawn at
half the maximum rate and represent estimates of the decision
boundaries.}
\label{fig1}
\end{figure}

\begin{figure}[t]
\centering
\includegraphics[width=85mm]{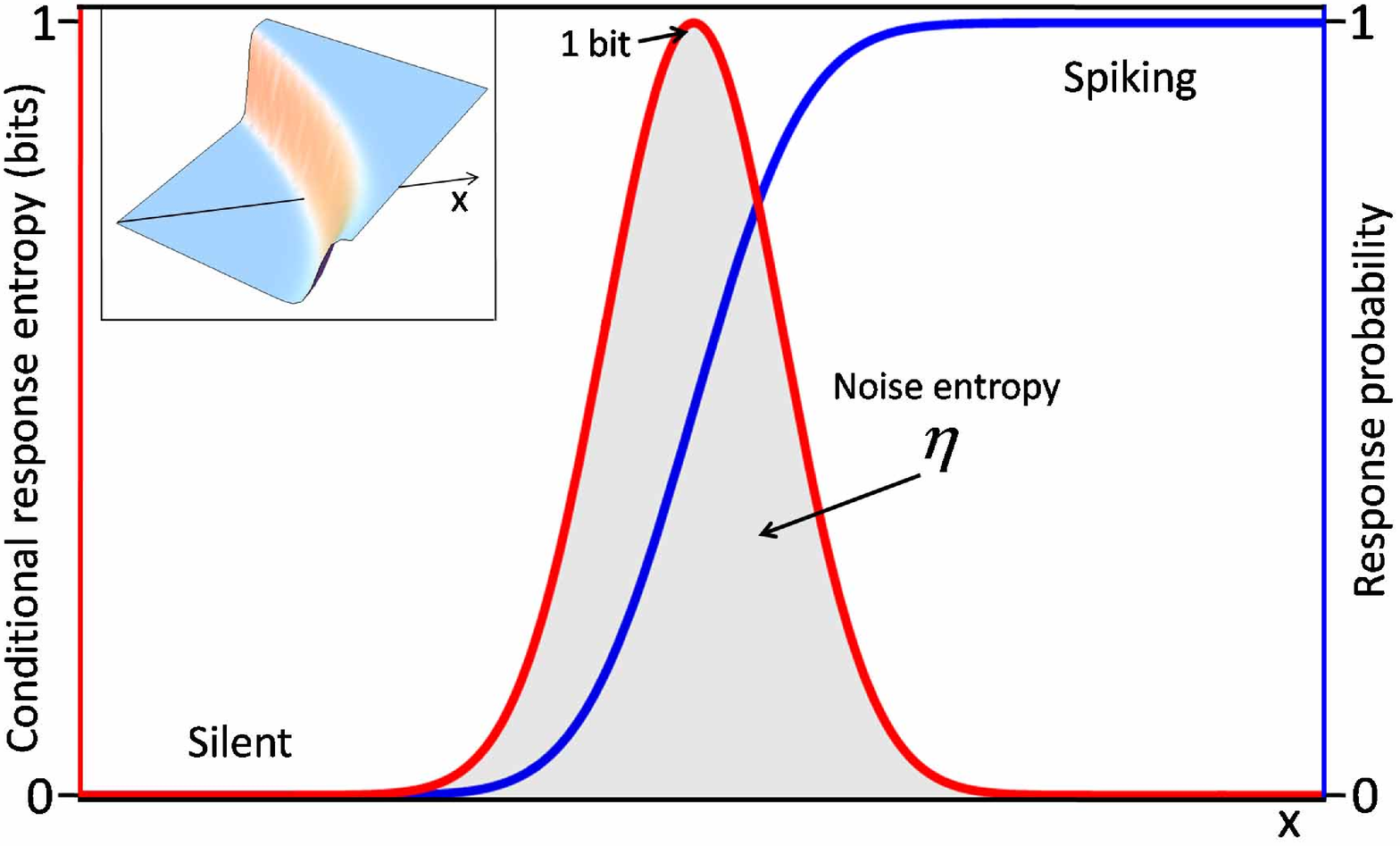}
\caption{(Color online) Schematic of response probability and noise entropy.  The response function in two dimensions (inset) is assumed to be deterministic everywhere except at the transition region which may curve in the input space.  In a direction x, perpendicular to some point on the decision boundary, the response function is sigmoidal (blue, no shading) going from silent to spiking.  The conditional response entropy (red, shading underneath) is $-P_{\rm{spike}|\rm{x}} \log_2 \left( P_{\rm{spike}|\rm{x}} \right)- \left( 1- P_{\rm{spike}|\rm{x}} \right)  \log_2 \left( 1- P_{\rm{spike}|\rm{x}} \right)$ and decays to zero at $x= \pm \infty $.  The contribution to the total noise entropy due to this cross-section, $\eta$, is the shaded area under the conditional response entropy curve.}
\label{noiseandresppic}
\end{figure}

In Fig. \ref{fig1} we provide examples of response functions estimated for two neurons in the cat primary visual cortex \cite{Sharpee06}.  For each neuron, the heat map shows the average firing rate in the space of the two most relevant input dimensions.  As this figure illustrates, even in two dimensions the mapping from inputs to the neural response (in this case the presence or absence of a spike in a small time bin) can be quite complex.  Nevertheless, one can delineate regions in the input space where the firing rate is above or below its average (red solid lines).  As an approximation, one can equate all firing rate values to the maximum value in regions where it is above average, and to zero in regions where it is below average.  This approximation of a sharp transition region of the response function is equivalent to assuming small noise in the response.  Across the boundary separating these regions, we will assume that the firing rate varies from zero to the maximum in a smooth manner (inset in Fig. \ref{noiseandresppic}).

As we discuss below, this approximation simplifies the response
functions enough to make the optimization problem tractable, yet it
still allows for a large diversity of nonlinear dependencies.  Upon
discretization into a binary variable, the firing rate of a single
neuron can be described by specifying regions in the input space where
spiking or silence is nearly always observed.  We will assume that
these deterministic regions are connected by sigmoidal transition
regions called \textit{decision boundaries}
\cite{Sharpee_Bialek_2007}, near which $P_{\sigma|\mathcal{I}} \approx
0.5$. The crucial component in the model is that the sigmoidal transitions
are sharp, affecting only a small portion of the input
space. Quantitatively, decision boundaries are well defined if the
width of the sigmoidal transition region is much smaller than
the radius of curvature of the boundary.

The decision boundary approach is amenable to the calculation of
mutual information.  The contribution to the noise entropy $H_{\textrm{noise}} $ from inputs near the boundary is on the order of one bit and decays to zero in the spiking/silent regions (Fig. \ref{noiseandresppic}). 
We introduce a weighting factor
$\eta$ to denote the summed contribution of inputs near a decision boundary obtained by
integrating $-\displaystyle\sum_{\sigma } P_{\sigma |x} \log_2 \left(
P_{\sigma |x} \right)$ \textit{across} the boundary. The factor $\eta$ depends on the
specific functional form of the transition from spiking to silence,
and represents a measure of neuronal noise. In a single-neuron system,
the total noise entropy is then an integral \textit{along} the
boundary, $H_{\rm{noise}} \approx \eta ~ \int_{\gamma} ds
~P_{\mathcal{I}} $, where $\gamma $ represents the boundary, and the
response entropy is $H_{\rm{resp}} = -P \log_2 P -\left( 1-P \right)
\log_2 \left( 1-P \right) $, where $P$ is the spike probability of the
neuron, equal to the integral of $P_{\mathcal{I}}$ over the spiking
region.

The decision boundary approach is also easily extended to the case of
multiple neurons, as shown in Fig. \ref{ndb_pic}(a).  In the
multi-neuronal circuit, the various response patterns are obtained
from intersections between decision boundaries of individual
neurons. In principle, all $2^{N}$ response patterns can be obtained
in this way. We denote as $G_{\sigma}$ the region of the space where
inputs elicit a response $\sigma$ from the network. To calculate the
response entropy for a given set of decision boundaries in a
$D-$dimensional space, the state probabilities are evaluated as
$D-$dimensional integrals over $\{G_\sigma\}$
\begin{equation}
\label{response prob}
P_{\sigma}=\displaystyle\int\limits_{G_{\sigma}}d^{D}\textbf{r}P_{\mathcal{I}}\left( \textbf{r} \right),
\end{equation}
weighted by $P_{\mathcal{I}}$. Just as in the case of a single neuron, the network response is assumed to be deterministic everywhere except near any of the transition regions. Near a decision boundary, the network can be with approximately equal probability in one of two states that differ in the response of the neuron associated with that boundary. Thus, such inputs contribute $\sim 1$ bit to the noise entropy, cf. Eq. (\ref{Hnoise}) and Fig. \ref{noiseandresppic}.  The total noise entropy can therefore be approximated as a surface integral over all decision boundaries weighted by $P_{\mathcal{I}}$,
\begin{equation}
\label{Hnoise_approx}
H_{\rm{noise}} \approx \eta \displaystyle\sum_{n=1}^{N} ~
\displaystyle\int\limits_{\gamma_{n}} ds ~P_{\mathcal{I}},
\end{equation}
where $\gamma_{n}$ is the decision boundary of the $n^{\text{th}}$
neuron. In this paper we will assume that $\eta$ is the same for all neurons and is
position-independent, but the extension to the more general case of
spatially varying $\eta$ is possible
\cite{Sharpee_Bialek_2007}. Finding the optimal mapping from inputs to
network states can now be turned into a variational problem with
respect to decision boundaries shapes and locations.

\section{Results}
\subsection{General solution for optimal coupling constants}
Our approach for finding the optimal coupling constants consists of
three steps. The first step is to find the optimal mapping from
inputs to network states, as described by decision boundaries. The
second step is to use this mapping to compute the optimal values of the 
response probabilities by averaging across all possible inputs. The
final step is to determine the coupling constants of the
Ising model from the set of optimal response probabilities.

Due to a high metabolic cost of spiking, we are interested in finding
the optimal mapping from inputs to network states that result in a
certain average spike probability across all neurons:
\begin{equation}
P = \dfrac{1}{N}\displaystyle\sum_{\sigma}S_\sigma P_{\sigma},
\label{Pdef}
\end{equation} 
where $S_\sigma$ is the number of ``up spins'', or firing neurons, in configuration $\sigma$. 
Taking metabolic constraints into account, we maximize the functional
\begin{eqnarray}
\label{functional}
F &=& H_{\rm{resp}}-H_{\rm{noise}}-\lambda \left( NP- \displaystyle\sum_{n=1}^{N} P_{n} \right) \nonumber \\
 & & -\displaystyle\sum_{\sigma} \beta_{\sigma} \left( P_{\sigma} -\displaystyle\int\limits_{G_{\sigma}}d^{D} \textbf{r} P_{\mathcal{I}} \left( \textbf{r} \right) \right)
\end{eqnarray}
\noindent where $ \lambda $, $ \left\lbrace \beta_{\sigma} \right\rbrace $ are Lagrange multipliers for the constraints and the last term demands self-consistency through Eq. (\ref{response prob}).

To accomplish the first step, we optimize the shape of each segment
between two intersection points. Requiring $\frac{\delta F}{\delta
\bf{r}}=0$ yields the following equation
\begin{eqnarray}
\eta\left[\kappa + \hat{\textbf{n}} \cdot \nabla P_{\mathcal{I}} \left( \textbf{r}\right) \right]+ \beta_{\sigma_{1},...,\sigma_{i-1},\uparrow,\sigma_{i+1},...,\sigma_{N}} \nonumber \\
-
\beta_{\sigma_{1},...,\sigma_{i-1},\downarrow,\sigma_{i+1},...,\sigma_{N}}=0
,
\label{equation1}
\end{eqnarray}
\noindent for the segment of the $i^{th}$ decision boundary that
separates the regions
$G_{\sigma_{1},...,\sigma_{i-1},\uparrow,\sigma_{i+1},...,\sigma_{N}}$
and
$G_{\sigma_{1},...,\sigma_{i-1},\downarrow,\sigma_{i+1},...,\sigma_{N}}$.
Here, $\hat{\textbf{n}}$ is the unit normal vector to the decision
boundary, and $\kappa = \nabla \cdot \hat{\textbf{n}} $ is the total
curvature of the boundary.  We then optimize with respect to the state
probabilities, $\frac{\delta F}{\delta P_{\sigma}}=0$, which gives a
set of equations
\begin{equation}
\beta_{\sigma}= S_{\sigma} \lambda -\dfrac{1}{\rm{ln}2} \left( 1 + \textrm{ln}  P_{\sigma} \right).
\label{equation2}
\end{equation}

\begin{figure}[t]
\centering \includegraphics[width=85mm]{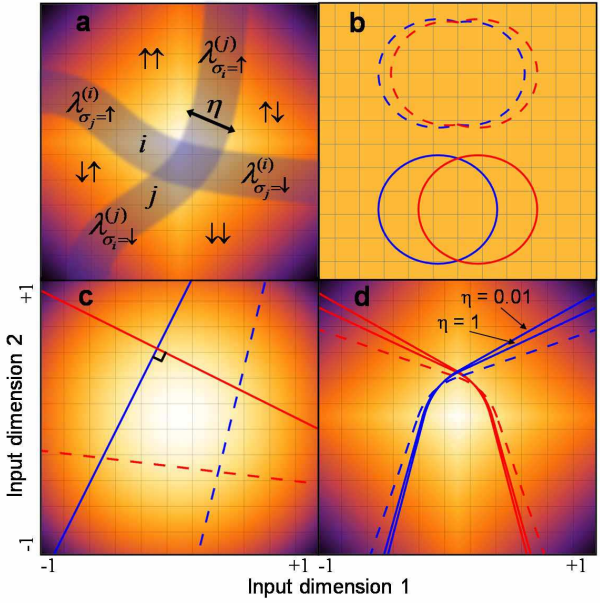}
\caption{(Color online) Network decision boundaries. Color in input space corresponds
to the input distribution $P_{\mathcal{I}}$ and each line is a
boundary for an individual neuron. \textbf{a}, For a general
intersection between neurons $i$ and $j$, the boundaries divide the
space into four regions corresponding to the possible network states
(e.g. $G_{\sigma_{i}=\uparrow, \sigma_{j}= \uparrow}$). Each segment
has a width $\eta$ which determines the noise level, and is described
by a parameter $\lambda^{\left( i\right)}_{\sigma'}$,
cf. Eq.~(\ref{segment}). \textbf{b}, In a uniform space, with two
neurons (different colors/shading), the boundary segments satisfying
the optimality condition are circular. Networks which have segments
with different curvatures (dashed) are less informative than smooth
circles (solid). \textbf{c}, In a Gaussian space (in units of standard
deviations), straight perpendicular lines (solid) provide more
information about inputs than decision boundaries that intersect at
any other angle (dashed). \textbf{d}, For approximately natural inputs
(plotted in units of standard deviations), the suboptimal balanced
solutions (dashed) are two independent boundaries with the same
$P$. The optimal solutions (solid) change their curvature at the
intersection point, and depend on the neuronal noise level $\eta$.}
\label{ndb_pic}
\end{figure}

Combining Eqs. (\ref{equation1}) and (\ref{equation2}), we arrive at
the following equation for the segment of the $i^{th}$ decision
boundary across which $\sigma_i$ changes while leaving the rest of the
network in state $\sigma'$:
\begin{equation}
\lambda^{\left( i \right) }_{\sigma'} + \kappa + \hat{\textbf{n}} \cdot \nabla P_{\mathcal{I}} \left( \textbf{r}\right) =0.
\label{segment}
\end{equation}
The parameter
\begin{equation}
\lambda^{\left( i \right) }_{\sigma'} = \dfrac{1}{\eta} \left[ \lambda  - \textrm{log}_{2} \left( \dfrac{P_{\sigma_{1},...,\sigma_{i-1},\uparrow,\sigma_{i+1},...,\sigma_{N}}}{P_{\sigma_{1},...,\sigma_{i-1},\downarrow,\sigma_{i+1},...,\sigma_{N}}} \right) \right],
\label{lambda}
\end{equation}
is specific to that segment, and is determined by the ratio of
probabilities of observing the states which this segment
separates. Generally, this ratio (and therefore the parameter
$\lambda^{\left( i \right) }_{\sigma'} $) may change when the
boundaries intersect.  For example, in the schematic in
Fig. \ref{ndb_pic}(a), $ \lambda^{\left( i \right)
}_{\sigma_{j}=\downarrow} $ depends on the ratio of $
P_{\sigma_{i}=\uparrow,\sigma_{j}=\downarrow } /
P_{\sigma_{i}=\downarrow,\sigma_{j}=\downarrow }$, whereas $
\lambda^{\left( i \right) }_{\sigma_{j}=\uparrow} $ depends on
$P_{\sigma_{i}=\uparrow,\sigma_{j}=\uparrow } /
P_{\sigma_{i}=\downarrow,\sigma_{j}=\uparrow } $.  The values of the
parameters for the two segments of the $i^{th}$ boundary are equal
only when $ P_{\sigma_{i}=\uparrow,\sigma_{j}=\downarrow } /
P_{\sigma_{i}=\downarrow,\sigma_{j}=\downarrow } =
P_{\sigma_{i}=\uparrow,\sigma_{j}=\uparrow } /
P_{\sigma_{i}=\downarrow,\sigma_{j}=\uparrow } $.  This condition is
satisfied when the neurons are independent. Therefore, we will refer
to the special case of a solution where $\lambda$ does not change its
value across any intersection points as an \textit{independent
boundary}. In fact, Eq. ({\ref{segment}}) with a constant $\lambda$ is the same as
was obtained in \cite{Sharpee_Bialek_2007} for a network with only one
neuron. In that case, the boundary was described by a single parameter
$\lambda$, which was determined from the neuron's firing rate. Thus, 
in the case of multiple neurons, the individual decision boundaries are
concatenations of segments of optimal boundaries computed for single
neurons with, in general different, constraints. A change in
$\lambda_{\sigma'}^{(i)}$ across an intersection point results in a
kink -- an abrupt change in the curvature of the boundary. Thus, by
measuring the change in curvature of a decision boundary of an
individual neuron one can obtain indirect measurements on the degree
of interdependence with other, possibly unmeasured, neurons.

Our main observation is that the $\lambda$-parameters determining
decision boundary segments can be directly related to the coupling
constants of the Ising model through a set of linear
relationships. For example, consider two neurons $i$ and $j$ within a
network of $N$ neurons whose decision boundaries intersect. It follows
from Eq. (\ref{lambda}) that the change in $\lambda$-parameters along
a decision boundary is the same for the $i$th and $j$th neuron, and is
given by:
\begin{eqnarray}
&& \lambda^{\left(i\right) }_{\sigma_{j}=\uparrow} 
- \lambda^{\left(i\right) }_{\sigma_{j}=\downarrow} =
\lambda^{\left(j\right)}_{\sigma_{i}=\uparrow} 
- \lambda^{\left(j\right)}_{\sigma_{i}=\downarrow} = \nonumber \\
&& = 
 \frac{1}{\eta} ~ \textrm{log}_{2} \left[ 
\frac{P_{\sigma^{\prime \prime},\sigma_{i}=\downarrow , \sigma_{j}= \downarrow}  P_{\sigma^{\prime \prime},\sigma_{i}=\uparrow , \sigma_{j}= \uparrow}  }
{P_{\sigma^{\prime \prime},\sigma_{i}=\uparrow , \sigma_{j}=
\downarrow } P_{\sigma^{\prime \prime},\sigma_{i}=\downarrow ,
\sigma_{j}= \uparrow} } \right].
\label{relation1a}
\end{eqnarray}
where $\sigma^{\prime \prime}$ represents the network state of all the neurons other than $i$ and $j$. 
Taking into account the Ising model via Eq.~(\ref{Ising Model Eq}),
this leads to a simple relationship for the interaction terms
$\left\lbrace J_{ij} \right\rbrace $:
\begin{equation}
\lambda^{\left(i\right)}_{\sigma_{j}=\uparrow} 
- \lambda^{\left(i\right)}_{\sigma_{j}=\downarrow} =
\lambda^{\left(j\right)}_{\sigma_{i}=\uparrow} 
- \lambda^{\left( j \right) }_{\sigma_{i}=\downarrow} 
= - \dfrac{2 \left( J_{ij}+J_{ji} \right)}{\eta  \ln 2} .
\label{relation1}
\end{equation}
We note that only the average (``symmetric'') component of pairwise
interactions can be determined in the Ising model. Indeed,
simultaneously increasing $J_{ij}$ and decreasing $J_{ji}$ by the same
amount will leave the Ising model probabilities unchanged because of
the perturbation symmetry in Eq. (\ref{Ising Model Eq}). This same limitation is present in the determination of $ \left\lbrace J_{ij} \right\rbrace $ via any method (e.g. an inverse Ising algorithm).

Once the interaction terms $ \left\lbrace  J_{ij} \right\rbrace  $ are known, the local
fields can be found as well from Eq.~(\ref{lambda}):
\begin{equation}
\label{relation2}
\lambda_{{\sigma}'}^{(i)}-\frac{\lambda}{\eta}=-\frac{2}{\eta  \ln 2 }\left[h_i+\sum_{k\neq i}J_{ik}\sigma_k\right].
\end{equation} 
This equation can be evaluated for any response pattern $\sigma'$,
because consistency between changes in $\lambda$ is guaranteed by
Eq.~(\ref{relation1}).

The linear relationships between the Ising model coupling constants
and the $\{\lambda^{\left( i \right) }_{\sigma'}
\}$ parameters are useful, because they can indicate what configurations of 
network decision boundaries can be consistent with an Ising
model. First, Eq. (\ref{relation1}) tells us that if the $i^{\rm{th}}$ boundary
is smooth at an intersection with another boundary $j$, then the
average pairwise interaction in the Ising model between neurons $i$
and $j$ is zero [as mentioned above the cases of truly zero
interaction $J_{ij}= J_{ji}=0$ and that of a balanced coupling
$J_{ij}= -J_{ji}$ cannot be distinguished in an Ising model].  Second,
we know that if one boundary is smooth at an intersection, then any
other boundary it intersects with is also smooth at that point. More
generally, the change in curvature has to be the same for the two
boundaries and we can use it to determine the average pairwise
interaction between the two neurons.  A third point is that the
change in curvature has to be the same at all points where the same
two boundaries intersect. For example, intersection between two planar
boundaries is allowed because the change is curvature is zero at
points across the intersection line. In cases where intersections form
disjoint sets the equal change in curvature would presumably have to
be due to a symmetry of decision boundaries.

In summary then, we have the analytical equations for the maximally
informative decision boundaries of a network, through
Eqs. (\ref{segment}) and (\ref{lambda}). We now study their solutions
for specific input distributions and then determine, through
Eqs. (\ref{relation1}) and (\ref{relation2}), the corresponding
maximally informative coupling constants.

\subsection{Uniform and Gaussian distributions}
We first consider the cases of uniformly and Gaussian distributed
input signals.  As discussed above, finding optimal configurations of
decision boundaries is a trade-off between maximizing the response
entropy and minimizing the noise entropy.  Segments of decision
boundaries described by Eq. (\ref{segment}) minimize the noise entropy
locally, whereas changes in the $\lambda$ parameters arise as a result
of maximizing the response entropy. The independent decision
boundaries, which have one constant $\lambda$ for each boundary,
minimize the noise entropy globally for a given firing rate.  Because
for one neuron, specifying the spike probability is sufficient to
determine the response entropy $P_{1}=P$, cf. Eq. (\ref{Pdef}), and
$H_{\textrm{resp}}=-P\log_2P- \left( 1-P \right) \log_2 \left( 1-P
\right) $, maximizing information for a given spike probability is
equivalent to minimizing the noise entropy
\cite{Sharpee_Bialek_2007}. When finding the optimal configuration of
boundaries in a network with an arbitrary number of nodes, the
response entropy is not fixed, because the response probability may
vary for each node (it is specified only on average across the
network). However, if there is some way of arranging a collection of
independent boundaries to obtain response probabilities that also
maximize the response entropy, then such a configuration must be
optimal because it simultaneously minimizes the noise entropy and
maximizes the response entropy. It turns out that such solutions are
possible for both the uniformly and Gaussian distributed input
signals.

For the simple case of two neurons receiving a two-dimensional
uniformly distributed input, as in Fig. \ref{ndb_pic}(b), the optimal
independent boundaries are circles, because they minimize the noise
entropy (circumference) for a given probability $P$ (area). In
general, the response entropy, Eq. (\ref{Hresponse}), is maximized for
the case of two neurons when the probability of both neurons spiking
is equal to $P^{2}$.  It is always possible to arrange two circular
boundaries to satisfy this requirement.  The same reasoning extends to
overlapping hyperspheres in higher dimensions, allowing one to
calculate the optimal network decision boundary configuration for
uniform inputs.  Therefore, for uniformly distributed inputs the
optimal network decision boundaries are overlapping circles in two
dimensions or hyperspheres in higher dimensions.

\begin{figure}[t]
\centering
\includegraphics[width=85mm]{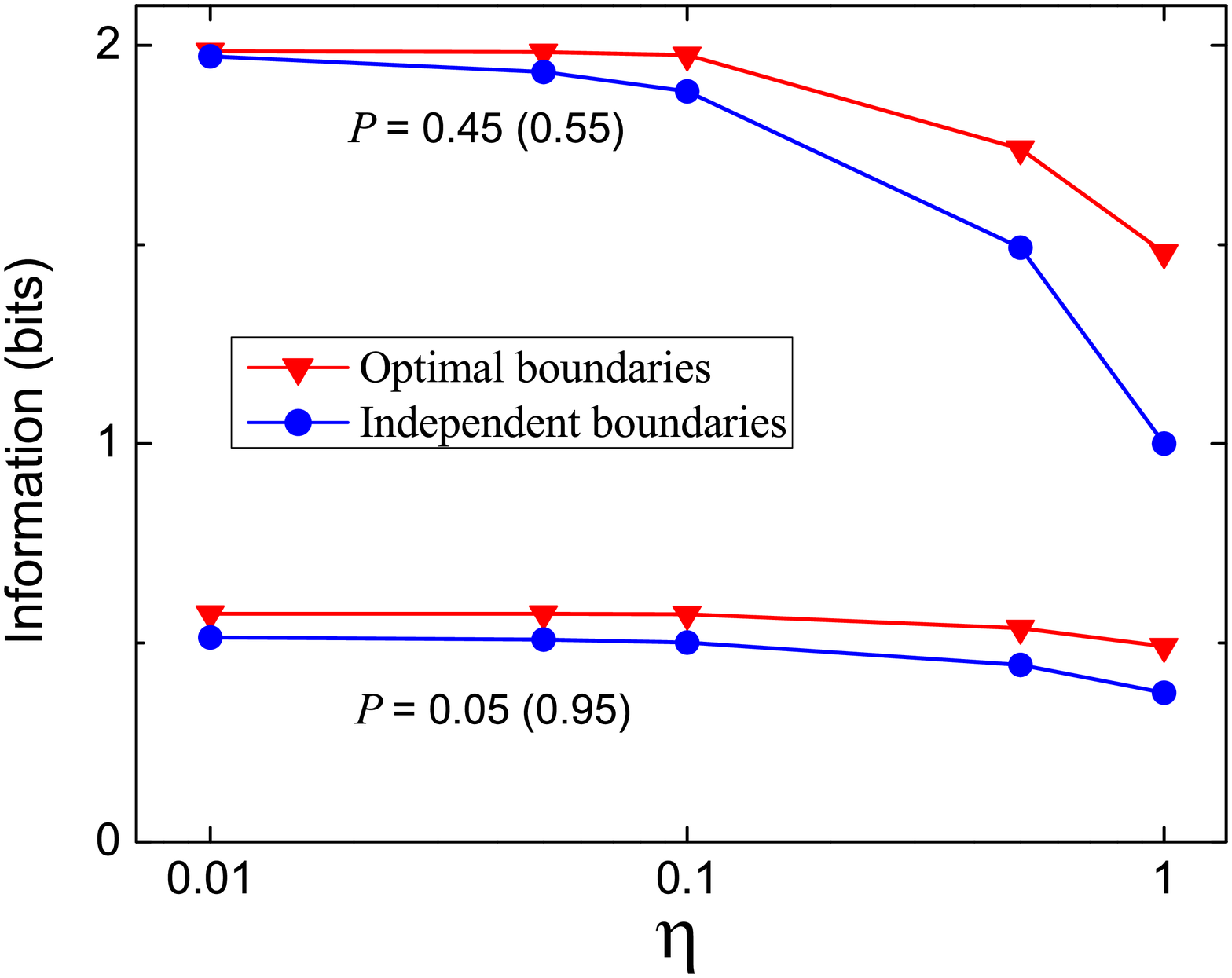}
\caption{(Color online) Information from independent
 and optimally coupled network boundaries as a function of  neuronal noise
 level. In both examples shown here, the independent boundaries (blue
 circles) lose information faster than the optimal boundaries (red
 triangles) as the noise level $\eta$ increases.  Each curve
 represents two response probabilities (e.g. $P=$ 0.45 and 0.55)
 because information is invariant under switching the spiking/silent
 regions. }
\label{info plot}
\end{figure}

For an uncorrelated Gaussian distribution, Fig. \ref{ndb_pic}(c), the
independent boundaries are $\left( D-1 \right) $-dimensional
hyperplanes \cite{Sharpee_Bialek_2007}. If we again consider two
neurons in a two-dimensional input space, then the individual response
probabilities for each boundary determine the perpendicular distance
from the origin to the lines. Any two straight lines will have the
same noise entropy, independent of the angle between them. However,
orthogonal lines (orthogonal hyperplanes in higher dimensions)
maximize the response entropy. The optimality of orthogonal boundaries
holds for any number of neurons and any input dimensionality.
We also find that for a given average firing rate across the network,
the maximal information per neuron does not depend on input
dimensionality as long as the number of neurons $N$ is less than the
input dimensionality $D$. For $N>D$, the information per neuron begins
to decrease with $N$, indicating redundancy between neurons.

\subsection{Naturalistic distributions}
Biological organisms rarely experience uniform or Gaussian input
distributions in their natural environments and might be
evolutionarily optimized to optimally process inputs with very
different statistics.  To approximate natural inputs, we use a
two-dimensional Laplace distribution, $ P_{\mathcal{I}} \left(
\textbf{r} \right) \propto \textrm{exp} \left( - \vert x \vert - \vert
y \vert \right) $, which captures the large-amplitude fluctuations
observed in natural stimuli \cite{Ruderman,Simoncelli01}, as well as
bursting in protein concentrations within a cell
\cite{Paulsson_2004}. For this input distribution there are four
families of solutions to Eq. (\ref{segment}) (see
\cite{Sharpee_Bialek_2007} for details), giving rise to many
potentially optimal network boundaries.  For a given $\lambda$, the
decision boundaries can be found analytically. To find the appropriate
value of the $\lambda$'s, we numerically solved Eq. (10) using
Mathematica \cite{mathematica6}. We found no solutions for independent
boundaries. The optimal boundaries therefore will have different
$\lambda$'s, and kinks at intersection points.  As a result, the
neurons will have a nonzero average coupling between them, examples of
which are shown in Fig. \ref{ndb_pic}(d).

We found that the shapes of the boundaries change with the noise level
$\eta$, which does not happen for independent boundaries
\cite{Sharpee_Bialek_2007}. To see if this noise dependence gives the network the ability
to compensate for noise in some way, we look at the maximum possible
information the optimal network boundaries are able to encode about
this particular input distribution for different noise levels
(Fig. \ref{info plot}).  We compare this to the suboptimal combination
of two independent boundaries with the same $P$. The figure
illustrates that the optimal solutions decrease in information less
quickly as the noise level increases.  The improvement in performance
results from their ability to change shape in order to compensate for
the increasing noise level.

\begin{figure}[t]
\centering
\includegraphics[width=85mm]{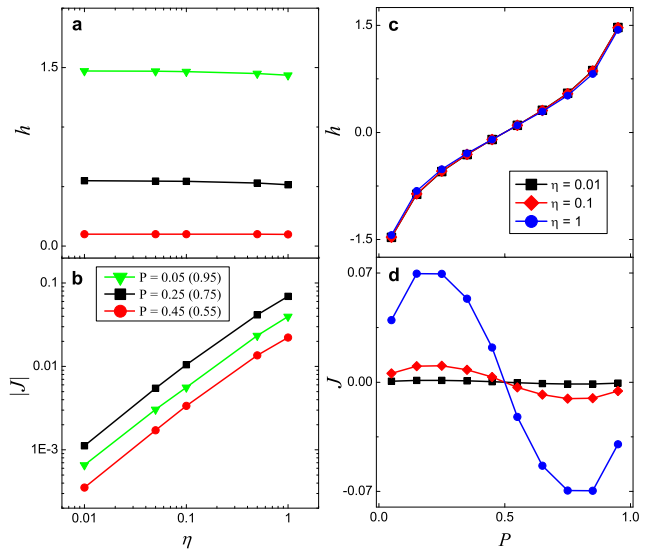}
\caption{(Color online) Optimal coupling constants
 for naturalistic inputs. \textbf{a}, The local fields show very
 little dependence on the noise level of the neurons, but the
 magnitude of the interaction strength $J$, \textbf{b}, increases in
 the same fashion with the noise level regardless of the response
 probability $P$.  \textbf{c}, $h$ depends strongly on the response
 probability. \textbf{d}, $J$ changes sign about $P=1/2$. Below this
 point the coupling is excitatory, and above it the coupling is
 inhibitory.} \label{couplingconstants}
\end{figure}

We calculated both $h\equiv h_{1} = h_{2}$ and $J\equiv \left(
J_{12}+J_{21} \right) / 2 $ for various noise levels and response
probabilities. Fig. \ref{couplingconstants}(a) and (c) show the local
field $h$ is practically independent of the noise level but does
depend on the response probability. The coupling strength, however,
depends on both noise level and response probability, increasing in
magnitude with neuronal noise, shown in
Fig. \ref{couplingconstants}(b). The combination of this result and
the noise compensation observed in Fig.~\ref{info plot}, suggests that
the network is able to use pairwise coupling for error correction,
with larger noise requiring larger coupling strengths. This strategy
is similar to the adage ``two heads are better than one'', to which we
add ``$\ldots$ especially if the two heads are noisy''.

In Fig. \ref{couplingconstants}(d) we observe that the sign of the
coupling changes as the value of $P$ crosses $1/2$. When $P=1/2$, the
optimal solution is an \textsf{X} crossing through the origin, which
is the only response probability for which the network boundary is
made of two independent boundaries, making $J=0$ for any noise level
when $P=1/2$. It can also be seen that $ J\rightarrow 0 $ as $ \eta
\rightarrow 0 $ for $P \neq 1/2$. Curiously, 
for a given $\eta$, the dependence of the optimal $J$ on $P$ is highly 
nonmonotonic: it changes sign across $P=1/2$, and reaches a maximum(minimum) value for $P\approx 0.25(0.75)$.

\section{Discussion}
The general mapping between inputs and maximally informative network
responses in the decision boundary approximation has allowed us to
calculate the Ising model coupling constants. In this approach,
network responses to a given input are not described by an Ising
model, which emerges after averaging network responses across many
different inputs. Although there are many configurations of network
decision boundaries that can be consistent with a pairwise-interaction
model, certain restrictions apply. For example, the change in
curvature of a decision boundary that can occur when two boundaries
intersect has to be the same for both boundaries, and, if they intersect 
at more than one disjoint surfaces then the curvatures of 
those surfaces must be the same.

We find that for both the uniform and Gaussian input distributions the
optimal network boundaries are independent. This implies that the
average interaction strength is zero for all pairs of nodes through
Eq. (\ref{relation1}).  Such balance between excitatory and inhibitory
connections has been observed in other contexts including rapid state
switching
\cite{Tsodyks_Sejnowski_1995}, Hebbian learning
\cite{Song_Abbott_2000}, and selective amplification
\cite{Murphy_Miller_2009}. In this context balanced coupling is just
one possible configuration of a network of decision boundaries and
happens to be optimal for uniformly and Gaussian distributed inputs.

For a more realistic input distribution, the Laplace distribution, we
found the optimal boundaries were not smooth at intersection points.
This indicates that the average coupling between the nodes in the
network should be non-zero to achieve maximal information
transmission. We also observed that the optimal configuration of the
network depended on the noise level in the responses of the nodes,
giving the network the ability to partially compensate for the
encoding errors introduced by the noise, which did not happen for the
less natural input distributions considered. Also, the fact that $J$
can be positive or negative between two nodes leads to the potential
for many stable states in the network, which could give the network
the capacity to function as autoassociative memory, as in the Hopfield
model \cite{Hopfield_1982,Amit_Sompolinsky_1985}. Similar network behaviors were reported in
\cite{Prentice_Balasubramanian_2009} for networks of ten neurons,
where the optimal coupling constants were numerically found for
correlated binary and Gaussian inputs. Our approach is different 
in that we use an Ising model to describe average network responses, but not responses to particular inputs, $P_{\sigma | \mathcal{I} }$.

Previous experiments have shown that simultaneous recordings from
neural populations could be well described by the Ising model.  In one
such experiment using natural inputs
\cite{Schneidman_2006}, the distributions of coupling constants showed
an average $h$ which was of order unity and negative and average $J$
which was small and positive.  Our results for the Laplace
distribution are in qualitative agreement with these previous findings
if one assumes a response probability $P<1/2$.  Due to the high
metabolic cost of spiking, this is a plausible assumption to make.  

The method we have put forth goes beyond predicting the maximally
informative coupling constants, to make statements about optimal
coding strategies for networks.  Although both uniform and Gaussian
inputs can be optimally encoded by balanced networks, for example,
their organizational strategies are remarkably different.  In the
uniform input case, the optimal boundaries curve in all dimensions,
meaning each node is attending to and encoding information about every
component of the possibly high-dimensional input, and they organize
themselves by determining the optimal amount of overlap between
boundaries. However, for the Gaussian distribution, each boundary is
planar, indicating that the nodes of the network are sensitive to only
one component of the input. The optimal strategy for networks
receiving this type of input is to attend to and encode orthogonal
directions in the input space, minimizing the redundancy in the coding
process. 

In terms of practical applications, perhaps the most useful aspect of
this framework is the ability to infer the strength of pairwise
interactions with other nodes in the network by examining decision
boundaries of single nodes, cf. Eq.~(\ref{relation1}).

The observation of different types of pairwise interactions for
networks processing Gaussian and naturalistic Laplace inputs raises
the possibility of discovering novel adaptive phenomena. Previous
studies in several sensory modalities have demonstrated that the
principle of information maximization can account well for changes in
the relevant input dimensions
\cite{Smirnakis97,Shapley78,kim_rieke,Srinivasan_1982,Buchsbaum_1983,Shapley79,Chander01,Hosoya05,Sharpee06,Theunissen00,Nagel_Doupe_2006,David04}
as well as the neural gain
\cite{Laughlin_1981,Brenner00adapt,Fairhall01,Nagel_Doupe_2006}
following changes in the input distribution. For example, nonlinear gain
functions have been shown to rescale with changes in input variance
\cite{Shapley79,Brenner00adapt,Fairhall01,Maravall_etal_2007,Nagel_2006}. Our results suggest that if 
neurons were adapted on one hand to Gaussian inputs, and then to
naturalistic non-Gaussian inputs, then multi-dimensional input/output
functions of individual neurons might change qualitatively, with
larger changes expected for noisier neurons.

By studying the geometry of interacting decision boundaries we have
gained insights into optimal coding strategies and coupling mechanisms
in networks.  Our work focused on the application to neural networks, 
but the method developed here is general to any network with nodes which 
have multidimensional, sigmoidal response functions.  
Although we have only considered three particular
distributions of inputs, the framework described here is general and
can be applied to other classes of inputs with the potential of
uncovering novel, metabolically efficient combinatorial coding
schemes.  In addition to making predictions for how optimal pairwise
interactions should change with adaptation to different statistics of
inputs, this approach provides a way to infer interactions with
unmeasured parts of the network simply by observing the geometric
properties of decisions boundaries of individual neurons.

\section*{Acknowledgments} 
The authors thank William Bialek and members of the CNL-T group for
helpful discussions.  This work was supported an Alfred P. Sloan
Fellowship, a Searle Scholarship, National Institute of Mental Health
grant No. K25MH068904, the Ray Thomas Edwards Career Development Award
in Biomedical Sciences, McKnight Scholar Award, Research Excellence
Award from W.M. Keck Foundation, and the Center for Theoretical
Biological Physics (NSF PHY-0822283).


\end{document}